\documentclass{aastex}

\usepackage{emulateapj5,epsfig,onecolfloat,apjfonts}   

\shorttitle{Cooling, pre-heating, and the Sunyaev--Zel'dovich effect}
\shortauthors{da Silva, Kay, Liddle, Thomas, Pearce \& Barbosa}

\submitted{}

\begin{document}

\twocolumn
[
\title{The impact of cooling and pre-heating on the Sunyaev--Zel'dovich effect}
\author{Antonio C. da Silva,$^1$ Scott T.~Kay,$^1$ Andrew R. Liddle,$^1$ Peter 
A. Thomas,$^1$\\ Frazer R. Pearce$^2$ and Domingos Barbosa$^{3}$}
\affil{$^1$Astronomy Centre, University of Sussex, Brighton BN1 9QJ, 
United Kingdom}
\affil{$^2$Physics and Astronomy, University of Nottingham, Nottingham NG7 2RD, 
United Kingdom}
\affil{$^3$Physics Division, Lawrence Berkeley Laboratory, University of 
California, Berkeley CA, U.~S.~A.}
\affil{$^4$CENTRA, Inst\'{\i}tuto Superior T\'{e}cnico, Lisboa, Portugal}
\begin{abstract}
We use hydrodynamical simulations to assess the impact of radiative cooling and 
`pre-heating' on predictions for the Sunyaev--Zel'dovich (SZ) effect. Cooling 
significantly reduces both the mean SZ signal and its angular power 
spectrum, while pre-heating can give a higher mean distortion while leaving the 
angular power spectrum below that found in a simulation without heating or 
cooling. We study the relative contribution from high and low density gas, and 
find that 
in the cooling model about 60 per cent of the mean thermal distortion arises 
from low overdensity gas.
We find that haloes dominate the 
thermal SZ power spectrum in all models, while in the cooling simulation the 
kinetic SZ power spectrum originates predominantly in lower overdensity gas.
\end{abstract}
]

\section{Introduction}

One of the most promising techniques for detecting overdense structures at high
redshift is from their imprint on the Cosmic Microwave Background (CMB), known
as the Sunyaev--Zel'dovich (SZ) effect (Sunyaev \& Zel'dovich 1972, 1980).  The
SZ effect arises due to scattering of the CMB photons from electrons in regions
of ionized gas.  It has two components, the thermal and kinetic effects (see
Birkinshaw 1999 for a review).  Successful SZ detections towards known
low-redshift galaxy clusters are now routine (e.g.~Jones et al.~1993; Myers et
al.~1997; Carlstrom et al.~2000) and observational studies are poised to move to
blank field surveys of the sky in an attempt to discover their high-redshift
counterparts (e.g.~Lo et al.~2000; Kneissl et al.~2001).  This has led to an
upsurge in theoretical work predicting what future surveys might see, the most
powerful technique being the use of $N$-body/hydrodynamical simulations to make
high-resolution maps (da Silva et al.~2000, 2001; Refregier et al.~2000; Seljak,
Burwell \& Pen 2001; Springel, White \& Hernquist 2001).  While these
simulations model gravitational collapse, adiabatic compression and
shock-heating of the baryons, they ignore the effects of radiative cooling which
affects the thermal state of the gas, particularly that in haloes (Pearce et
al.~2000; Muanwong et al.~2001) where the SZ signal is strongest.  Additionally,
it is now believed that the baryons were subjected to non-gravitational heating
processes, probably before clusters formed (e.g.~Ponman, Cannon \& Navarro
1999); such `pre-heating' could come from the feedback of energy from supernovae
or AGN, and is therefore inherently related to the galaxy formation process.

In this Letter, we present a first attempt to simulate the SZ effect 
using hydrodynamical simulations which implement models for both
radiative cooling and  pre-heating. While these
simulations cannot be regarded as definitive, they illustrate how these 
processes affect predictions for
SZ-related quantities. We focus on the simplest measurable quantities,
which are the mean thermal distortion, and the dispersion and angular
power spectra of the SZ-induced temperature anisotropies. We also
quantify the relative contributions to the SZ effect from bound gas in
haloes and the low-overdensity intergalactic medium (IGM). 

~

\section{Simulation details and SZ map-making}

We present results from three simulations of the currently favoured 
$\Lambda$CDM cosmology. We set the cosmological parameters as follows:
matter density, $\Omega_{{\rm m}}=0.35$; cosmological constant, 
$\Omega_{\Lambda}\equiv\Lambda/3H_0^2=0.65$; Hubble constant,
$h=H_0/100 \, {\rm km\,s^{-1}}$ ${\rm Mpc^{-1}}=0.71$;
baryon density, $\Omega_{{\rm B}} h^2=0.019$; normalization
$\sigma_8=0.9$. The density field was
constructed using $N=4,096,000$ particles of both baryonic and dark matter, 
perturbed from a regular grid of fixed comoving size 
$L=100 \, h^{-1} {\rm Mpc}$. The dark matter and baryon particle
masses are $2.1\times 10^{10} \, h^{-1} {\rm M_{\odot}}$ and
$2.6 \times 10^{9} \, h^{-1} {\rm M_{\odot}}$ respectively. In physical
units, the gravitational softening was held fixed to $25\,h^{-1} {\rm
kpc}$ below $z=1$ and above this redshift scaled as $50(1+z)^{-1}\,h^{-1} {\rm 
kpc}$.
Each simulation took approximately 2300 time steps to evolve to $z=0$,
using a parallel version of the {\sc hydra} code (Couchman, Thomas \& Pearce 
1995; Pearce \& Couchman 1997), which uses the 
adaptive particle-particle/particle-mesh (AP$^3$M) algorithm 
to calculate gravitational forces (Couchman 1991) and
smoothed particle hydrodynamics (SPH) to estimate hydrodynamical
quantities. Our SPH implementation follows that used by 
Thacker \& Couchman (2000), where more details can be found.

The first of the three simulations was performed without any additional
heating or cooling mechanisms; we refer to this as the `non-radiative'
simulation. The second simulation included a model for radiative cooling
using the method described in Thomas \& Couchman (1992), except that we adopted
the cooling tables of Sutherland \& Dopita (1993)
and a global gas metallicity
evolving as $Z=0.3(t/t_0) Z_{\odot}$, where $t_0 \simeq 12.8 \, {\rm Gyr}$ 
is the current age of the Universe. Cooled gas is converted into 
collisionless material and no longer participates in generating the SZ
effect. This fraction is negligible at $z=6.5$
(the highest redshift we consider in computing the SZ effect) but
increases to $\sim 20$ per cent by the present. Although this
model provides acceptable matches to the $z=0$ X-ray group/cluster 
scaling relations (Muanwong et al.~2001), the fraction of cooled 
material is uncomfortably high when compared to observational determinations
(e.g.~Balogh et al.~2001). Hence, this simulation provides
an extreme case in terms of the amount of cooling obtained.

Our third simulation, which also includes cooling, was performed in order to 
study the effects of pre-heating
the gas at high redshift.  Although attempts are now being made to include
effective energy feedback from supernovae within cosmological simulations
(e.g.~Kay et al.~2001), the modelling can only be done at a phenomenological
level since the true physical processes occur on much smaller scales
than are resolved.  For the purposes of this paper, we adopt a particularly
simple model, which is to take the output from the radiative simulation at $z=4$
and raise the specific thermal energy of all gas particles by 0.1 keV (an
equivalent temperature of $1.2\times 10^{6}\,$K), before evolving to $z=0$ as
before.  This choice of energy is motivated by observations of an `entropy
floor' in X-ray groups by Lloyd-Davies, Ponman \& Cannon (2000); we define
specific `entropy' as $s=k_BT/n^{2/3}$ and our injection
model would give $s_{\rm floor} \sim 80 \, {\rm keV \, cm^2}$ were the gas all
at mean density.  This reduces the amount of cooled material to a more
acceptable level ($\sim 10$\%), although a greater amount of heating would have 
been needed to give the best match to the X-ray scaling relations.

These simulations have two advantages over those we studied previously
(da Silva et al.~2000, 2001). Firstly, those simulations did not include
radiative cooling, and so contain hot high-density gas where the cooling time 
is much shorter than a Hubble time. Secondly, the new simulations 
have improved numerical resolution (the particle masses are a factor of 19
smaller) allowing the SZ effect to be studied at smaller scales. 

We analyze the SZ effect by constructing simulated SZ
maps of area one square degree,\footnote{A collection of colour images and 
animations can be viewed at
{\tt www.astronomy.sussex.ac.uk/users/antonio/sz.html}} stacking simulation 
boxes from low to high redshift (up to redshift $z \sim 6.5$) as
described by da Silva et al.~(2000).  The vast majority of the
thermal SZ signal is produced well within this redshift interval. 
We repeat this process to produce 30 random map realizations for each
run, using the same sequence of random seeds for the different models. The 
thermal effect is described by the Comptonization parameter $y$, and the kinetic 
by the temperature perturbation $\Delta T/T$.

\section{Results}

\subsection{The mean thermal distortion and the rms dispersion}

The mean $y$ distortion, obtained by averaging $y$ over lines of sight, is an
important measure of the global state of the gas in the Universe. In our 
simulations we find
\begin{tabbing}
~~~~ \= Non-radiative:~~~~ \= $y_{{\rm mean}} = 3.2 \times 10^{-6}$ \\
\> Cooling: \> $y_{{\rm mean}} = 2.3 \times 10^{-6}$ \\
\> Pre-heating: \> $y_{{\rm mean}} = 4.6 \times 10^{-6}$. 
\end{tabbing}
(For comparison, the mean mass-weighted temperature at $z=0$ in the three 
simulations are 
0.39, 0.35
and 0.38 keV.)  At present, the best upper limit is $y_{{\rm mean}} < 1.5 \times
10^{-5}$ from {\it COBE--FIRAS} (Fixsen et al.~1996), and all our models are
well below this.  The cooling simulation gives the lowest level of distortion
and the pre-heating the highest.  The factor of two difference between these
runs shows that the mean $y$ is quite sensitive to pre-heating.  The mean $y$
distortion in the non-radiative model is slightly smaller than the value found
in da Silva et al.~(2000); part of this difference is attributable to a 
different $\sigma_8$, and the remainder lies within the uncertainty expected 
from changes to the simulation technique.

\begin{figure}[t]
\begin{center}
\epsfxsize=8.70cm \epsfbox{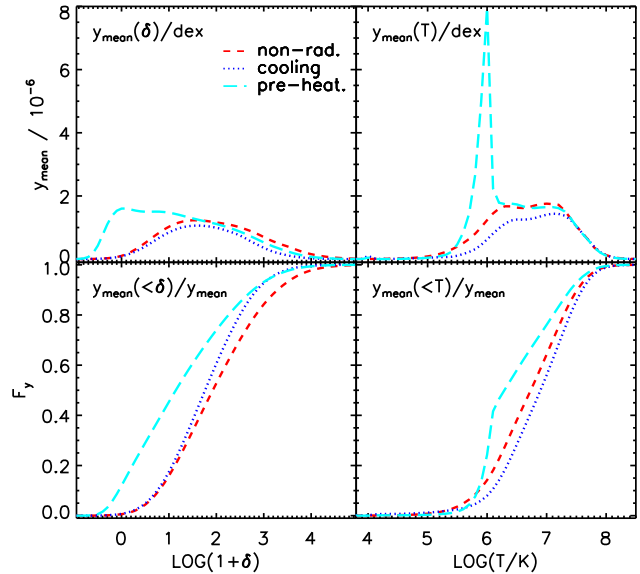}
\figcaption{\label{f:ytd} Fractional contribution to $y_{{\rm
mean}}$ as a function of temperature and overdensity. The upper panels show the 
contribution for a unit interval (in base ten logarithm), while the lower show 
the integrated contribution.} 
\end{center}
\end{figure}

To study the relative contribution to the SZ effect from the different gas
phases in the simulations, we sorted the particles in each simulation according
to their temperature, $T$, and overdensity, $\delta$, and computed the fraction
of SZ signal arising from all particles at a given $T$ and $\delta$.  By
carrying out cuts in the $\delta$--$T$ plane we are able to determine the type
of structures contributing to the SZ effect.  In Figure~\ref{f:ytd}, the top
panels show the distribution of the mean $y$ as a function of overdensity and
temperature.  The area under the curves are the $y_{{\rm mean}}$ listed earlier.
The bottom panels show the corresponding cumulative mean $y$ fractions,
$F_{y}=y_{{\rm mean}}(<\!\delta)/y_{{\rm mean}}$ and $F_{y}=y_{{\rm
mean}}(<\!T)/y_{{\rm mean}}$.

The left-hand panels show that the mean $y$ receives contributions
from a wide range of overdensities. Cooling reduces the contribution from
high-density gas in groups and clusters. Because most of the gas in
the simulations is at low overdensity, pre-heating preferentially
raises the contribution from that phase.
The right-hand panels in Figure~\ref{f:ytd} indicate that gas below $10^5$K and 
above $10^8$K contributes negligibly to $y_{{\rm mean}}$.  
The mean $y$ distributions have similar shapes in the cooling and
non-radiative models and are practically uniform in the range
$6.2\lesssim \log (T/{\rm K})\lesssim 7.3$ for all runs. Gas above $\sim 2
\times 10^7$K produces the same amount of distortion independently of
the model considered.
The pre-heating run shows a sharp peak at $T\simeq
10^6$K resulting from IGM gas heated to this
temperature by the energy injection at $z=4$.
While pre-heating will indeed insert a new temperature scale, the
narrowness of this feature is no doubt an artifact of our simplified
method and would be broadened in reality. Indeed, there is evidence that some of 
the IGM is cooler than this (Schaye et al.~2000).

\begin{figure}[t]
\begin{center}
\epsfxsize=8.70cm \epsfbox{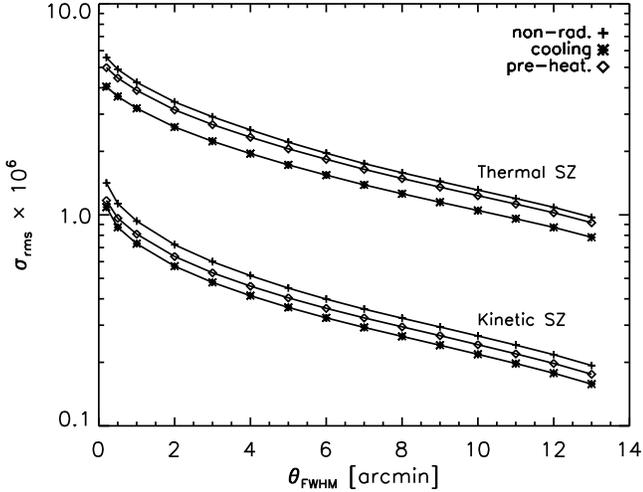}
\figcaption{\label{f:sigrms} Dispersion $\sigma_{{\rm rms}}$ 
for the thermal and kinetic effects, as a function of beam 
resolution (assuming gaussian beams with FWHM=$\theta _{\rm FWHM}$). }
\end{center}
\end{figure}

Figure~\ref{f:sigrms} shows the rms temperature perturbation $\sigma_{{\rm 
rms}}$ as a function of beam resolution $\theta_{{\rm FWHM}}$, obtained by
averaging the 30 thermal (Rayleigh--Jeans band) and kinetic SZ maps.  The 
different cooling/pre-heating models shift the
curves by tens of percent.  On the angular scales shown, the kinetic dispersions
are a factor $\sim$5 smaller than the thermal, confirming earlier results using
lower-resolution simulations (da Silva et al.~2001).


\begin{figure}[t]
\begin{center}
\epsfxsize=8.70cm \epsfbox{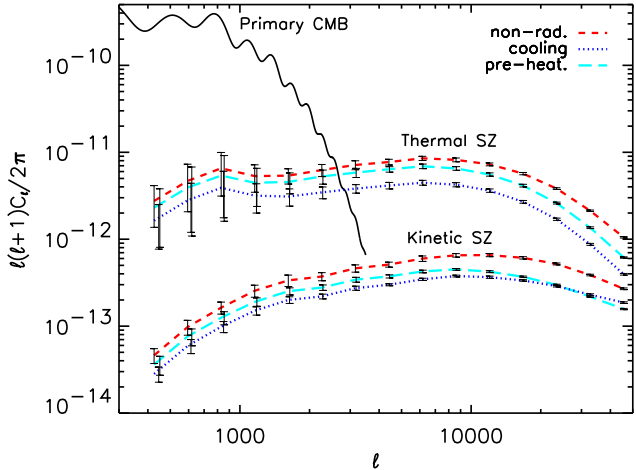}
\figcaption{\label{f:cls} Thermal and kinetic angular
power spectra from the three simulations. Curves are averages over 30
maps and bars are $1\sigma $ bootstrap errors.}
\end{center}
\end{figure}

\subsection{The angular power spectra}

Figure~\ref{f:cls} shows the SZ angular power spectra in the Rayleigh--Jeans
region, with $C_\ell$ defined in the usual way.  In each case, the curves show
averages over 30 sky realizations, and the error bars are statistical errors at
68 per cent confidence level obtained by bootstrap resampling the set of maps.
The map-to-map variations are typically much higher than these and in
particular, as we will see below, the large angular scales are significantly
affected by sample variance as well as the finite map size.  Comparing the
different models, we see that the curves have similar shapes, with the thermal
power spectra dominating the kinetic ones on all scales, and dominating the
primary CMB anisotropies for $\ell>3000$.  Radiative cooling has the effect of
lowering the SZ spectra by a factor of about two as compared with the
non-radiative case.  Our model of pre-heating raises the spectra again, though
only part way back towards those found in the non-radiative run.

\begin{figure}[t]
\begin{center}
\epsfxsize=8.70cm \epsfbox{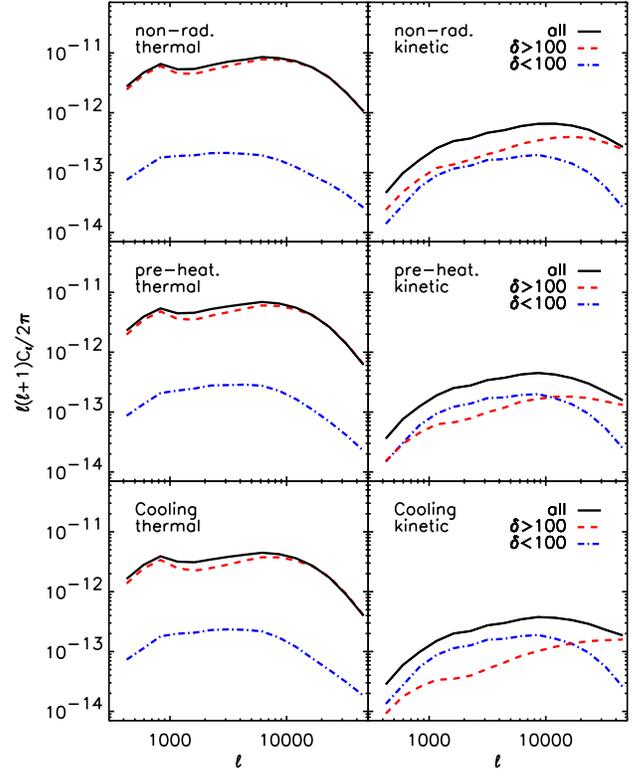}
\figcaption{\label{f:cls_phases} 
Contribution to the spectra from gas in haloes
($\delta >100$) and the low-overdensity intergalactic medium ($\delta <100$).}
\end{center}
\end{figure}

Figure~\ref{f:cls_phases} shows the contributions to the power spectra from the
low-overdensity IGM ($\delta<100$) and from bound gas in haloes ($\delta>100$).
Most of the thermal signal comes from the bound gas, with the IGM contribution
an order of magnitude below.  The kinetic signal is dominated by halos in the
non-radiative run, by the IGM phase in the cooling simulation, and has roughly
equal contributions from each in the pre-heating model.  Accordingly, treatments
of the kinetic effect which concentrate solely on bound haloes are likely to
miss much of the kinetic signal.  For both the thermal and kinetic effects, the
contribution from low-density gas is similar in all three models, and the
difference in the spectral amplitudes is therefore mainly due to the
contribution of the high-density gas.  These differences mainly reflect the
different amounts of cooled material in haloes in each simulation.  At $z=0$ the
amount of cooled material in the pre-heating run is a factor of about two less
than in the cooling simulation (while the non-radiative model does not include
star formation at all).

\begin{figure}[t]
\begin{center}
\epsfxsize=8.70cm \epsfbox{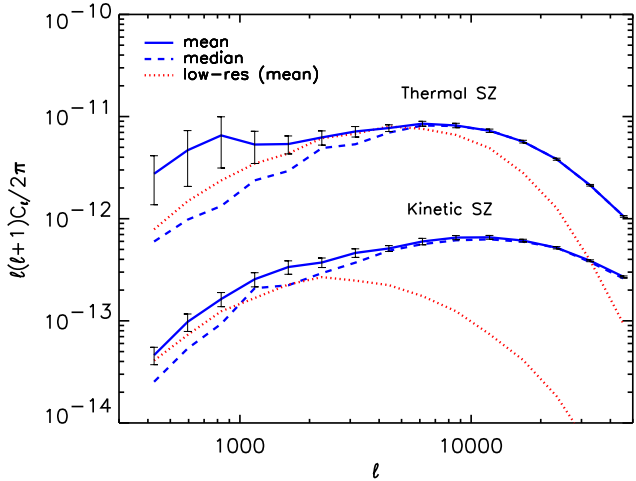}
\figcaption{\label{f:cls2} Effects of sample variance and resolution on the SZ 
spectra obtained from non-radiative simulations. The curves are the mean power 
spectra (solid lines) and the median of the $C_l$ distributions
(dashed lines). The mean power spectra from da Silva et al.~(2001)
are shown as the dotted lines.}
\end{center}
\end{figure}

We end by discussing how cosmic variance and numerical resolution affects these
determinations. Even though our spectra are averages over 30 maps, there
remains some sensitivity to rare features, and because these are nongaussian
they can have large effects, as reflected in the bootstrap errors.  One of the 
30 thermal maps is dominated by a very nearby bright source, strong enough to 
greatly boost the estimated power spectrum. In Figure~\ref{f:cls2} the solid 
line reproduces the mean power spectrum from Figure~\ref{f:cls}, while the 
dashed line shows the median power spectrum, which is close to what would have 
been obtained had a single map with a bright source by chance not appeared in 
the set. The median is consistent with the bootstrap errors (which of course 
include subsamples without that map); note however that it is the mean, not the 
median, which is the maximum likelihood power spectrum estimator. We conclude 
that
there is considerable sample variance error in estimating the thermal power 
spectrum from observations with limited survey areas.  The dotted lines show the 
SZ power
spectra obtained from the non-radiative $\Lambda$CDM simulation studied in da
Silva et al.~(2001), which had a mass resolution 19 times worse than the present
simulations. The loss of power due to resolution in the earlier results is
clearly seen at high $\ell$, an effect particularly severe in the kinetic case, 
and our new results extend the $\ell$-range for which the power spectra are 
reliably obtained. Based on the demonstrated convergence of the low-resolution 
simulations, we estimate the thermal power spectrum is accurately determined out 
to at least $\ell$ of 20,000.

Our non-radiative SZ angular power spectra show broad agreement with
the results obtained by other authors, who use only non-radiative
hydrodynamical simulations. After accounting for the differing
$\sigma_8$ and $\Omega_{{\rm B}} h$, on large scales our mean thermal spectrum
agrees well with the results of Refregier et al.~(2000) and Refregier \& 
Teyssier (2000), whereas
the predictions from Seljak et al.~(2000) and Springel et al.~(2000) are closer 
to our median $C_\ell$s. On small scales our mean thermal spectrum
shows somewhat less power and a similar shape to the results in
Springel et al.~(2000) and Refregier \& Teyssier (2000), while our kinetic 
angular
power spectrum is in good agreement with that of Springel et al.~(2000).

\section{Conclusions}

We have analyzed the impact on the magnitude and spectrum of the SZ effect of 
physical processes in the hot gas, including both radiative cooling and a crude 
model for pre-heating of the IGM. While neither radically alters the appearance 
of the SZ sky, the details of how the signal is generated are modified. Cooling 
reduces the angular power spectrum by a factor of around two as compared to 
non-radiative simulations, while our pre-heating model gives a significant boost 
to the mean distortion but a more minor change to the power spectra.

We have studied these effects in more detail by classifying the gas into two 
phases. 
While the mean thermal distortion receives equal contribution from gas in haloes 
and the IGM, the thermal angular power spectrum is dominated by contributions 
from hot gravitationally-bound gas.
By contrast, the kinetic power spectrum receives equal contributions from 
both phases in the pre-heating run and is dominated by the IGM in the cooling 
model. For both the thermal and kinetic SZ spectra, the contribution from IGM
gas is similar in all three models, whereas the contribution from gas in haloes
is largest in the non-radiative simulation and lowest in the cooling run.

\acknowledgments

ACdS and DB were supported by FCT (Portugal), and STK, PAT and FRP by PPARC 
(UK). The work presented in this paper was carried out as part of the 
programme of the Virgo Supercomputing Consortium ({\tt 
www.virgo.sussex.ac.uk}).

\end{document}